\documentclass[runningheads]{llncs}

\usepackage{graphicx}
\usepackage{floatrow}
\usepackage{fancyhdr}
\usepackage{listings}
\usepackage{graphicx}
\usepackage{amssymb}
\usepackage{colortbl,hhline}
\usepackage[misc]{ifsym} 

\usepackage[inline]{enumitem} 
\usepackage{subcaption}       

\begin{document}

\title{ROC: An Ontology for Country Responses towards COVID-19}

\author{
Jamal Al Qundus\inst{}\orcidID{0000-0002-8848-1632}\and 
Ralph Sch\"afermeier\inst{*}\orcidID{0000-0002-4349-6726}\and 
Naouel Karam\inst{}\orcidID{0000-0002-6762-6417}\and
Silvio Peikert\inst{}\orcidID{0000-0001-5716-1540}\and 
Adrian Paschke\inst{}\orcidID{0000-0003-3156-9040}
}

\authorrunning{Al Qundus et al.}

\institute{
Fraunhofer FOKUS, Kaiserin-Augusta-Allee 31, 10589 Berlin, Germany
\email{ralph.schaefermeier@fokus.fraunhofer.de}
}

\maketitle 

\begin{abstract}
The ROC ontology for country responses to COVID-19 provides a model for collecting, linking and sharing data on the COVID-19 pandemic. It follows semantic standardization (W3C standards RDF, OWL, SPARQL) for the representation of concepts and creation of vocabularies. ROC focuses on country measures and enables the integration of data from heterogeneous data sources. The proposed ontology is intended to facilitate statistical analysis to study and evaluate the effectiveness and side effects of government responses to COVID-19 in different countries. The ontology contains data collected by OxCGRT from publicly available information. This data has been compiled from information provided by ECDC for most countries, as well as from various repositories used to collect data on COVID-19.

\keywords{Ontology Engineering \and Semantic Web Technologies \and Covid-19 Ontology \and Novel Coronavirus Ontology \and Covid-19 Responses \and Pandemic Country Responses}
\end{abstract}

\thispagestyle{fancy}
\fancyhf{}
\fancyhead[R]{}
\fancyfoot[C]{\tiny Copyright \textcopyright\ 2021 for this paper by its authors.\\ Use permitted under Creative Commons License Attribution 4.0 International (CC BY 4.0).}
\renewcommand{\headrulewidth}{0pt}

\setcounter{footnote}{0}

\section{Introduction}
\label{sec:introduction}


Almost all countries are affected by the (second wave) COVID-19 pandemic. Many organisations have setup systems and projects to collect and publish global data on the impact of the virus allowing to monitor the pandemic. Most countries have implemented responsive measures as reaction to COVID-19. The impact of the pandemic and policies implemented differ widely among countries. Meanwhile data on the infection spread and government responses during the first phase of the pandemic (March to July 2020) is available to enable research on the effects of individual policies. 

Policy makers face tough decisions on how to deal with the pandemic, since responses considered most effective have also significant negative impacts  on economy and social life. This is strongly reflected in the measures implemented by the countries. The policies implemented and the reactions of society vary greatly among countries and cities. While many people support their government strategy, others reject measures included. Empirical studies need to be conducted to support decision-makers and information campaigns. Which responses have proven to be effective in containing COVID-19 and what are their effects on the economy and society? are questions that apply to individual countries as well as to cultural, political and geographical regions.

To enable empirical research, global data on infections, recoveries, death rates and policies implemented in various countries and regions is published via various sources. The challenge for data scientists and epidemiologists is that these data sources do not share common standards or methodologies for reporting their data. The reported data is influenced by time zones/holidays, different political and/or economic incentives, variation in counting methods\footnote{cause and place of death -at home or in hospital-}, discovered / undiscovered numbers, etc. In addition, in many cases country's reports differ depending on the reporting institute. For example, the numbers of infected people in Germany differs between Johns Hopkins Coronavirus Resource Center\footnote{\url{https://coronavirus.jhu.edu/map.html}} (USA) and Robert Koch institute\footnote{\url{https://rki.de/DE/Content/InfAZ/N/Neuartiges\_Coronavirus/Fallzahlen.html}} (GER). Unfortunately, this difference is not comprehensible. To study this valuable information and perform statistical analysis on it, a common standard for harmonizing and reporting the data is required.

Many researchers are taking up this challenge, and the first solutions to this problem have already been developed. The COviD-19 ontology (CODO) is an ontology for organizing cases data and patient information and aims to use the technology of knowledge graphs to analyse the pandemic \cite{dutta2020codo}. Our work follows the same approach to ontology design and has a common motivation. Nevertheless, it focuses on areas not yet covered, government responses to the pandemic, and therefore this work is equally unique in this field. The ontology developed in this paper addresses the following goals:
\begin{itemize}
\item[$\bullet$] serve as a reference scheme for use in reporting COVID-19 related data.
\item[$\bullet$] provide a common conceptualization and thereby abstract from heterogeneous structures of existing sources of (static) data and provide a common linking schema, which is essential for data accessibility.
\item[$\bullet$] provide a defined data structure with a fixed semantics for further analysis or monitoring systems.
\item[$\bullet$] offer a supplement of the existing ontologies in order to build a common global data model.
\item[$\bullet$] offer a template to organize data from other pandemics or nationwide events or measures.
\item[$\bullet$] (as a follow-up outcome) enable countries to make coordinated strategic decisions while avoiding lockdowns and their entailed risks
\end{itemize}

For the ontology development process, we followed a hybrid approach, which is driven by available data on the one hand and questions of interest that should be answerable by the ontology on the other hand. We mainly collected data from OxCGRT\footnote{\url{https://www.bsg.ox.ac.uk/}}, ILO\footnote{\url{https://www.ilo.org/global/lang--en/index.htm}} and ECDC\footnote{\url{https://www.ecdc.europa.eu/}}. ECDC provides data on infection rates, OxCGRT systematically monitors government actions related to the pandemic and ILO provides economic indicators focused on the labour market. The modeling of the data consisted of the following steps: manual review and merging of the data. Concepts were extracted and linked using logical relationships. To design and create the ontology, we used Protégé and then ingested the data based on the ROC ontology.

The remainder of the paper is organized as follows:
Section~\ref{sec:relatedwork} contains a review on related work. Section~\ref{sec:methodology} describes the development methodology. Section~\ref{sec:therocontology} introduces the ROC ontology and its concepts. Section~\ref{sec:rocevaluation} describes the data ingestion process and querying capabilities. Section~\ref{sec:summary} concludes the paper with a short summary and future work.

\section{Related Work}
\label{sec:relatedwork}

This section provides an overview on related work investigating ontologies related to diseases, especially relevant to COVID-19.

The Human Disease Ontology (DO)\footnote{\url{http://www.disease-ontology.org}} classifies thousands of human diseases moving to a multi-editor model in web ontology language enabling collaboration of several working groups, and has recently been extended by DOID\footnote{\url{https://disease-ontology.org/term/DOID:11725/}} including COVID-19 concepts \cite{10.1093/nar/gku1011}. The Infectious Disease Ontology (IDO) \cite{babcock2020infectious} and the Ontology of Coronavirus Infectious Disease (CIDO) \cite{he2020cido} define vocabulary relating to infectious diseases such as flu, malaria, and brucellosis. CIDO additionally represents comparative analysis of COVID-19 and other diseases wrt. symptoms, drugs, clinical trials etc.

One of the most relevant work is the Ontology for cases and patient information (CODO)\footnote{\url{https://github.com/biswanathdutta/CODO}}, which initiated the development of CIDO. It provides a standards-based and comprehensive open source model for data collection on the COVID 19 pandemic. The ontology is very well suited for the integration of data from heterogeneous data sources and thus represents one of the major inspirations for our work. Furthermore, our methodology follows the procedure for term definition described in the work. This is based on the reuse of concepts from other leading vocabularies and the use of the W3C standards RDF, OWL, SWRL and SPARQL. The evaluation of CODO was conducted on the basis of data received from the Indian government \cite{dutta2020codo}.

Of course, the ontologies (COVIDCRFRAPID)\footnote{\url{https://bioportal.bioontology.org/ontologies/COVIDCRF
RAPID}} of the World Health Organization (WHO) as data model for the case COVID-19 RAPID, the ontologies Kg-COVID-19\footnote{\url{https://github.com/Knowledge-Graph-Hub/kg-covid-19}} for the creation of knowledge graphs including SARS-COV-2 and the ontology Linked-Data COVID-19\footnote{\url{https://zenodo.org/record/3765375\#.XraWJmgzbIU}} are also relevant.  These ontologies contain conceptualizations (and partially instance data) related to COVID-19 cases and are primarily aimed at software applications such as question \& answering or monitoring dashboards \cite{dutta2020codo}. However, to our knowledge, the work described in this paper is the first work to take into account government responses and link them to epidemiolical data, such as case data.

All these works support the containment of the pandemic, but more importantly, the works build on and complement each other. The focus so far has been mainly on gathering cases, symptoms and information from infected people, for example to identify and isolate hot spots. However, one field has not been covered so far, namely the field of countries government responses. It is precisely this knowledge gap that our work with the ROC ontology aims to fill.

In the next sections we describe the methodology for the development of the ROC ontology based on the state-of-the-art principles, its structure and evaluation.

\section{Methodology}
\label{sec:methodology}

The artifact created and investigated in this work is an ontology authored using the Web Ontology Language~2 (OWL~2)\footnote{\url{https://www.w3.org/TR/owl2-overview/}}, the semantics of which are based on description logics.

The main purpose of the ontology, as pointed out in Section \ref{sec:introduction}, is the integration of public data on national responses to the COVID-19 pandemic and to provide a layer of interoperability between different and diverse resources as well as to answer interesting questions from the data.
The development of the ontology was therefore bottom-up data-driven as well as top-down application driven.
A further requirement was the integration of existing ontologies in the domain of COVID-19 in order to avoid redundancy as well as to leverage knowledge gained from the integration of existing knowledge (such as the combination of data about national responses with case data).

A multitude of ontology development methods exist, comprising, but not limited to METHONTOLOGY \cite{FLG+97} and On-To-Knowledge \cite{sure:2004aa}, both of which provide general ontology development guidelines and principles, Diligent \cite{tempich:2007aa}, which defines an argumentation-based development process as well as agile development methods, such as RapidOWL \cite{Auer:fj}.

The selection of the ontology development method was driven by the non-functional requirements to
\begin{enumerate*}[label=(\alph*\upshape)]
    \item \label{item:req-int-data} integrate existing data sources and
    \item \label{item:req-int-onto} integrate existing ontological sources
\end{enumerate*}
as well as 
\begin{enumerate*}[resume,label=(\alph*\upshape)]
    \item \label{item:req-content} by content-related requirements, which are formulated in the form of competency questions.
\end{enumerate*}

We decided to use the NeOn methodology \cite{DBLP:books/daglib/p/Suarez-FigueroaGF12}, since it  provides guidelines for either of the above-mentioned requirement types and allows combining them.
NeOn identifies a set of scenarios and provides an ontology development method for each scenario.
The scenarios applicable to the development of the ROC ontology are scenario 1 ``From specification to implementation'' (which involves requirement engineering and is suitable for our non-functional requirement \ref{item:req-content}), scenario 2 ``Reusing and re-engineering non-ontological resources'' , as we use external, non-ontological data sources (item \ref{item:req-int-data}) and scenario 3 ``Reusing ontological resources'', since we reuse concepts from the CODO ontology and align concepts derived from the external data sources with existing ones in the CODO ontology manually (item \ref{item:req-int-onto}).

The ontology development process therefore consists of three main activities, which start in parallel and which are divided into subtasks, some of which interact (see Figure \ref{fig:methodology}):

\begin{figure}
    \centering
    \includegraphics[width=.6\textwidth]{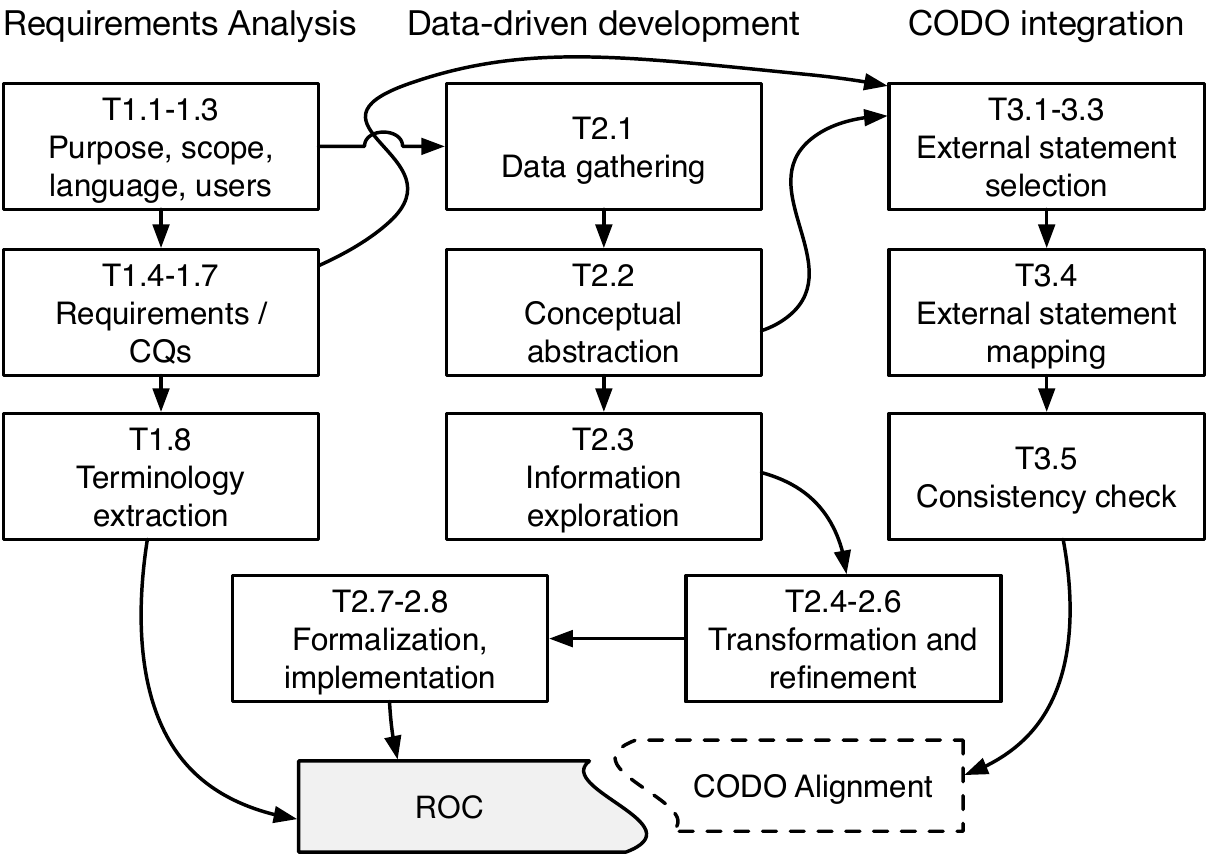}
    \caption{The development process of the ROC ontology}
    \label{fig:methodology}
\end{figure}

Activities 2 and 3 resulted in the ROC ontology, while activity 3 resulted in a set of import and mapping axioms (which are also part of the ROC ontology).

\section{The ROC Ontology}
\label{sec:therocontology}

The ontology\footnote{The current version 1.0 of the ROC ontology is publicly available at \\ \url{http://qurator-csi.de/ontologies/roc}} was designed around a set of competency questions, for instance, one can ask the following:

\begin{enumerate}[label=\textbf{CQ\arabic*}]
    \item Which countries do establish a certain response?
    \item At which incidence level do individual countries establish certain responses / response levels?
    \item How long do individual countries keep their response measures active?
    \item Were countries which established responses at low incidence levels able to avoid high incidence rates?
    \item Is there an effect of certain responses on infection rates? Can we measure effect or delay of effect? 
\end{enumerate}

The ontology consists of 27 OWL classes, 10 object properties, 42 data properties and 3 annotation properties.
Its expressivity is $\mathcal{ALEHI(D)}$, i.e., $\mathcal{AL}$ with full existential qualification, role hierarchy, inverse roles and data types, making it a member of the OWL 2 DL profile. The central domain concepts of the ROC ontology are the indicators as defined by the Oxford Covid-19 Government Response Tracker (OxCGRT) mentioned in Section \ref{sec:introduction}.

\begin{figure}
    \centering
    \begin{subfigure}{.69\textwidth}
        \includegraphics[scale=.31]{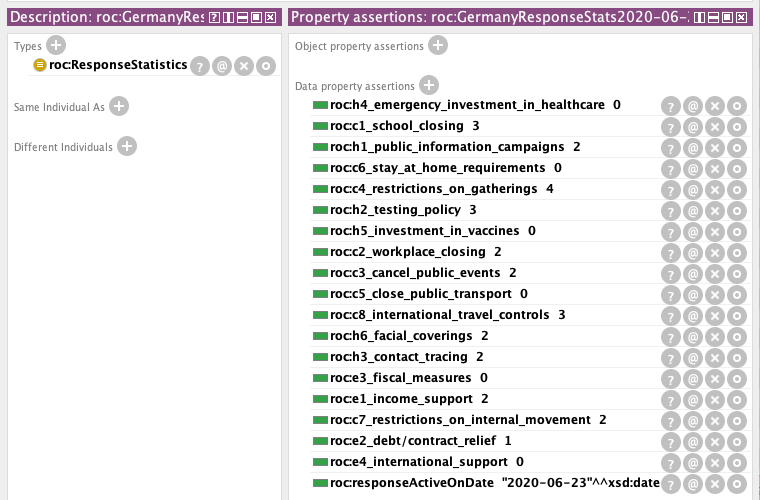}
        \caption{A response statistics instance with acquired response values}
        \label{fig:response-inst}
    \end{subfigure}
    \begin{subfigure}{.3\textwidth}
        \centering
        \includegraphics[scale=.31]{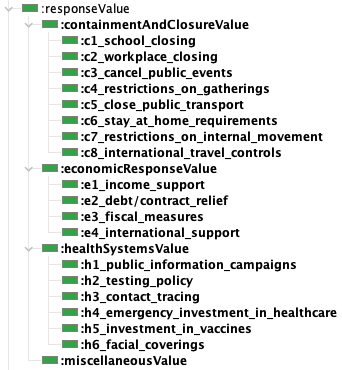}
        \caption{The hierarchy of data properties representing the OxCGRT response values}
        \label{fig:response-dp}
    \end{subfigure}
\end{figure}

The values for each of these indicators is acquired and stored in a data record.
A record is represented in the ROC ontology as an instance of the class \textsf{ResponseStatistics}, which is a subclass of the CODO class \textsf{CountryWiseStatistics} (see Figure \ref{fig:response-inst}).

In alignment with the structure of the data sources, which assign numerical values to each of these indicators, we modeled these indicators as data properties.
We established a data property hierarchy reflecting the taxonomy of OxCGRT coding categories (containment and closure (C), economic response (E), health systems (H), and miscellaneous (M)) (see Figure \ref{fig:response-dp}).
Creating common super properties for each category of response values allows for a Description Logic reasoner to infer that if any of the response values in a certain category has a value, then the super data property (representing the whole category) has a value.

\section{ROC in use}
\label{sec:rocevaluation}
This section describes how the ontology is used to ingest a set of data collected from multiple sources. It outlines the data transformation process and how to query the resulting RDF knowledge base (KB) to answer competency questions.

\subsection{Data ingestion}
\begin{figure*}[h!]
\centering
  \includegraphics[width=1\textwidth]{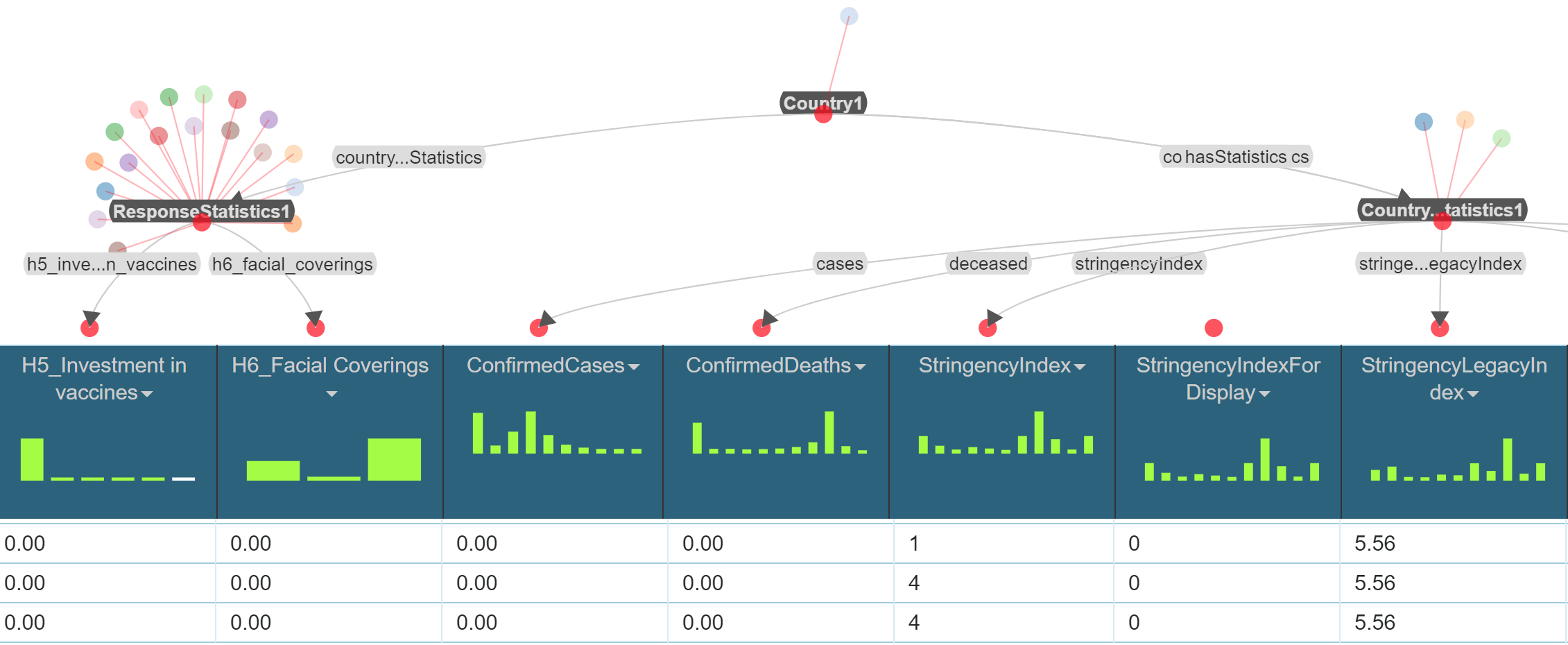}
\caption{A peek at the data mapping and transformation interface}
\label{fig:karmaMapping}
\end{figure*}

We manually reviewed and merged data collected from OxCGRT, ILO and ECDC. We then transformed data coming from those different sources into RDF based on the ROC ontology. The resulting KB serves as an integrated view to answer queries spreading over all required sources.
For data transformation, we made use of the Karma integration tool \cite{taheriyan2016websem}. The tool offers a user interface (as depicted in Figure \ref{fig:karmaMapping}) for mapping different types of structured data and publish it in an RDF format. It automatically generates an R2RML\footnote{\url{https://www.w3.org/TR/r2rml/}} mapping model based on users input. The model can be stored and reused on similarly structured data.

The resulting KB contains 1850 instances with data for Germany, Jordan and Sweden collected between January and the beginning of November 2020. We loaded the RDF data into the Virtuoso\footnote{\url{https://virtuoso.openlinksw.com/}} triple store, making it accessible and queryable through the SPARQL endpoint.

\subsection{Data querying}
We devised an initial set of queries to answer the competency questions over the RDF data. As an example, for the question "List the countries and their respective health responses?" the corresponding query would be the one depicted in Listing ~\ref{lst:sparql}.

\counterwithout{lstlisting}{chapter}

\begin{lstlisting}[ captionpos=b, caption=SPARQL query for question:"List the countries and their respective health responses?", label=lst:sparql, basicstyle=\scriptsize\ttfamily,frame=single]
PREFIX roc: <http://qurator-csi.de/ontologies/covid/responses#>
PREFIX codo: <http://www.isibang.ac.in/ns/codo#>
SELECT ?country AVG(?testing_policy) AVG(?contact_tracing)
SUM(?investment_healthcare)  SUM(?investment_in_vaccines)
AVG(?facial_coverings)
WHERE {
    ?country codo:countryWiseStatistics ?stats.
    ?stats roc:h2_testing_policy ?testing_policy;
    roc:h3_contact_tracing ?contact_tracing;
    roc:h4_emergency_investment_in_healthcare ?investment_healthcare;
    roc:h5_investment_in_vaccines ?investment_in_vaccines;
    roc:h6_facial_coverings ?facial_coverings.
}
GROUP BY ?country
\end{lstlisting}

\begin{figure*}[h!]
\centering
  \includegraphics[width=1\textwidth]{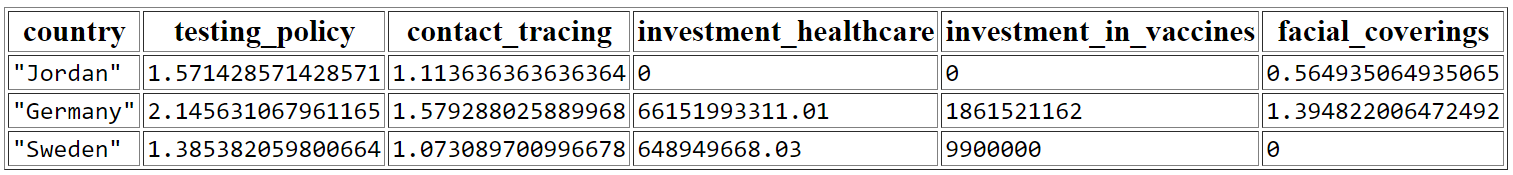}
\caption{SPARQL query results}
\label{fig:results}
\end{figure*}

The query results are depicted in Figure~\ref{fig:results}. We can see for instance that Germany has a higher testing policy index and the highest emergency investments in  healthcare and vaccines. We can also note that Sweden did not implement a facial covering policy. Based on the correlation between countries stringency and actual statistics, the effectiveness or failure of implemented measures can be derived, informing government future decisions.

\section{Summary and Next Steps}
\label{sec:summary}
The present work focuses on country responses against COVID-19 and proposes a novel ontology ROC to enable the integration of data from heterogeneous data sources and answer interesting questions. This facilitates statistical analysis to investigate and evaluate the effectiveness and side effects of such responses. The ontology consists of 27 OWL classes, 10 object properties, 42 data properties and 3 annotation properties. The data collected by OxCGRT, ILO and ECDC were manually reviewed and merged. Then we converted data from these different sources into RDF based on the ROC ontology. The resulting RDF serves as an integrated view to answer queries that span all required sources. The resulting KB contains 1850 instances of data for Germany, Jordan and Sweden collected between January and early November 2020. We uploaded the RDF data to the Virtuoso Triple Store and made it accessible and queryable through a SPARQL endpoint.

Given the fact that most experts are not familiar with SPARQL nor with Semantic Web technologies, we plan to connect our Controlled Natural Language querying system \cite{KaramSKCP20}, offering a user-friendly interface for querying the KB.

Other factors could influence the effectiveness of country responses. These are either difficult to capture\footnote{such as culture, acceptance of people, discipline, people's perseverance, family structure, types of contacts in society, etc.} or relatively easy to determine\footnote{such as public events during the pandemic, foreign trade, tourism, weather, country's experience of pandemics, time (how long the reactions will last), government system, structure, infrastructure, relationship between people and government (trust), etc}. Consideration of such factors would lead to the extension of the properties of the concepts/terms in order to increase the semantic expressiveness of the ontology.

Lastly, a technical and a goal-based evaluation of the effectiveness of the approach is a challenge to be addressed by future work.

\section*{Acknowledgements}
The research presented in this article is partially funded by the German Federal Ministry of Education and Research (BMBF) through the project QURATOR (Unternehmen Region, Wachstumskern, grant no.~03WKDA1F). \url{http://qurator.ai}

\bibliographystyle{splncs04}
\bibliography{bibliography}

\end{document}